\documentclass{article}

     \PassOptionsToPackage{round,sort&compress}{natbib}

\usepackage[final]{neurips_2024}

\usepackage[utf8]{inputenc} %
\usepackage[T1]{fontenc}    %
\usepackage[pdftex,bookmarksnumbered,bookmarksopen,colorlinks,citecolor={blue!70!black},linkcolor={blue!70!black},urlcolor=blue]{hyperref}
\usepackage{url}            %
\usepackage{booktabs}       %
\usepackage{amsfonts}       %
\usepackage{nicefrac}       %
\usepackage{microtype}      %
\usepackage{graphicx}
\usepackage[capitalize]{cleveref}
\usepackage[table]{xcolor}
\usepackage{tikz}
\usepackage[group-separator={,},group-minimum-digits={3}]{siunitx}

\usepackage{amssymb}  %
\usepackage{enumitem} %
\usepackage{pifont}   %

\renewcommand{\cite}{\citep}
\newcommand{\yep}{$\bullet$}

\usepackage{amssymb}

\title{
Participatory Assessment of Large Language Model Applications in an Academic Medical Center

}

\usepackage{fontawesome5}

\usepackage{authblk}

\newcommand{\chuv}{1}
\newcommand{\unil}{2}

\author[\chuv]{\bf Giorgia Carra}
\author[\chuv,\unil]{\bf Bogdan Kulynych}
\author[\chuv]{\bf François Bastardot}
\author[\chuv]{\bf Daniel Kaufmann}
\author[\chuv]{\\\bf Noémie Boillat-Blanco}
\author[\chuv,\unil]{\bf Jean Louis Raisaro}
\affil[\chuv]{Lausanne University Hospital, Switzerland}
\affil[\unil]{University of Lausanne, Switzerland}

\begin{document}

\maketitle

\begin{abstract}

Although Large Language Models (LLMs) have shown promising performance in healthcare-related applications, their deployment in the medical domain poses unique challenges of ethical, regulatory, and technical nature. In this study, we employ a systematic participatory approach to investigate the needs and expectations regarding clinical applications of LLMs at Lausanne University Hospital, an academic medical center in Switzerland. Having identified potential LLM use-cases in collaboration with thirty stakeholders, including clinical staff across 11 departments as well nursing and patient representatives, we assess the current feasibility of these use-cases taking into account the regulatory frameworks, data protection regulation, bias, hallucinations, and deployment constraints. This study provides a framework for a participatory approach to identifying institutional needs with respect to introducing advanced technologies into healthcare practice, and a realistic analysis of the technology readiness level of LLMs for medical applications, highlighting the issues that would need to be overcome LLMs in healthcare to be ethical, and regulatory compliant.

\end{abstract}

\section{Introduction}
Large Language Models (LLMs) represent a significant step forward in computer systems' ability to process and work with text-based information. The healthcare field, which routinely handles a large amount of written documentation, sees LLMs as a useful tool to help overcome pressing issues~\cite{thirunavukarasu2023large,clusmann2023future} such as lack of qualified clinical professionals, alarm fatigue, an ever-growing administrative burden, and  increasing pressure due to an aging society. 
Indeed, LLMs have shown promising performance in several healthcare tasks including medical writing, editing, summarization~\cite{madden2023summarymedicalnotes,umeton2024gpt,wu2024epfl}.

In this study, we use a systematic participatory approach to evaluate the emerging needs and expectations regarding LLM applications in a European academic medical center---Lausanne University Hospital in Switzerland---with a representative pool of thirty healthcare professionals across different roles and hospital departments, and discuss technical, ethical, and regulatory feasibility of these use-cases. Our analysis shows what technical advances, institutional changes, and regulatory updates would be needed to implement these systems effectively.

\begin{table}[t]
\newcommand{\rankcolor}[2]{%
  \ifnum#1=1\textcolor[HTML]{472877}{#2}\else%
  \ifnum#1=2\textcolor[HTML]{3E4989}{#2}\else%
  \ifnum#1=3\textcolor[HTML]{30688E}{#2}\else%
  \ifnum#1=4\textcolor[HTML]{25838E}{#2}\else%
  \ifnum#1=5\textcolor[HTML]{1E9E89}{#2}\else%
  \ifnum#1=6\textcolor[HTML]{35B778}{#2}\else%
  \ifnum#1=7\textcolor[HTML]{6ECE58}{#2}\else%
  \ifnum#1=8\textcolor[HTML]{B5DE2B}{#2}\else%
  \textcolor[HTML]{000000}{#2}\fi\fi\fi\fi\fi\fi\fi\fi%
}
\setlength{\arrayrulewidth}{0.5pt}
\arrayrulecolor{lightgray}
\centering
\vspace{1em}
\caption{\textbf{Average ranking of LLM applications by a working group of healthcare professionals at the Lausanne University Hospital.} We show the ranks according to six criteria regarding the impact of the applications and their perceived safety, along with the total (summed) ranks. Lower numbers are better.}
\label{tab:use-cases}
\resizebox{\linewidth}{!}{
\begin{tabular}{l *6{w{c}{.5cm}|} c}
 & \multicolumn{6}{c|}{Impact Criteria} & \multicolumn{1}{c}{} \\
\textbf{Application} & \textbf{Inst} & \textbf{QC} & \textbf{PE} & \textbf{WL} & \textbf{ME} & \textbf{Saf} & \textbf{Total} \\
\cmidrule(l{3pt}r{3pt}){1-1} \cmidrule(l{3pt}r{3pt}){2-7} \cmidrule(l{3pt}r{3pt}){8-8}
Automatic tool for classifying incident reports & \rankcolor{5}{5} & \rankcolor{2}{2} & \rankcolor{3}{3} & \rankcolor{6}{6} & \rankcolor{2}{2} & \rankcolor{1}{1} & \rankcolor{1}{19} \\
\rowcolor[HTML]{F7F6F6}
Summarization of documentation into patient-adapted language & \rankcolor{4}{4} & \rankcolor{5}{5} & \rankcolor{2}{2} & \rankcolor{3}{3} & \rankcolor{3}{3} & \rankcolor{2}{2} & \rankcolor{1}{19} \\
Automatic generation of draft discharge letters & \rankcolor{1}{1} & \rankcolor{7}{7} & \rankcolor{6}{6} & \rankcolor{1}{1} & \rankcolor{1}{1} & \rankcolor{6}{6} & \rankcolor{3}{22} \\
\rowcolor[HTML]{F7F6F6}
Assistive chatbot for clinical decision-making & \rankcolor{2}{2} & \rankcolor{1}{1} & \rankcolor{5}{5} & \rankcolor{4}{4} & \rankcolor{4}{4} & \rankcolor{7}{7} & \rankcolor{4}{23} \\
Entry notes and patient history summarization & \rankcolor{3}{3} & \rankcolor{6}{6} & \rankcolor{4}{4} & \rankcolor{2}{2} & \rankcolor{5}{5} & \rankcolor{4}{4} & \rankcolor{5}{24} \\
\rowcolor[HTML]{F7F6F6}
Analysis of patients' feedback and complaints & \rankcolor{7}{7} & \rankcolor{4}{4} & \rankcolor{1}{1} & \rankcolor{5}{5} & \rankcolor{6}{6} & \rankcolor{3}{3} & \rankcolor{5}{26} \\
Chatbot for medical education, guidelines retrieval, and literature summarization & \rankcolor{6}{6} & \rankcolor{3}{3} & \rankcolor{7}{7} & \rankcolor{7}{7} & \rankcolor{7}{7} & \rankcolor{5}{5} & \rankcolor{6}{35} \\
\end{tabular}
}
\\[1em]
\scriptsize
\textbf{Inst}: Institutional, \ 
\textbf{QC}: Quality of care, \
\textbf{PE}: Patient engagement, \ 
\textbf{WL}: Workload, \ 
\textbf{ME}: Measurability, \
\textbf{Saf}: Safety
\end{table}

\section{Methods}

We organized a working group (WG) of institutional stakeholders to discuss the needs, opportunities and strategic priorities regarding LLMs.
The working group included 30 participants from across the hospital: clinical staff from 11 departments, nursing and medical directors, patient advocates, legal experts, IT staff, clinical informaticians, and representativies of the Biomedical Data Science Center of the hospital.
We held regular monthly discussions among the members over a six-month period.

\paragraph{Preliminary discussions} The initial meeting focused on discussing the vision, brainstorming institutional needs, and identifying potential applications where LLMs could add value. In the second meeting, we established key evaluation criteria to score the use-cases. Evaluation criteria were mostly based on transversal goals such as potential impact on patient engagement, quality of care, reduced administrative workload.

\paragraph{Phase I: Survey}  Following the identification of a broad list of use-cases and their evaluation criteria, we designed and distributed a structured survey among the WG members. The survey aimed to assess the use-cases which seemed most promising to the WG members, as well as obtain the initial assessments of the application in terms of the identified criteria. Based on the results of the survey, we identified seven most voted applications, which we studied further in the next phase. See \cref{app:survey-details} for details.

\paragraph{Phase II: Participatory workshop} 
We held a participatory workshop in which the members of the WG organizing team presented in detail each of the seven use-cases identified as most popular in the survey, and facilitated a discussion among the members on various aspects of the use-cases. After the discussions, we conducted an anonymous survey in which the members ranked each of the applications according to the criteria established previously.

\section{Results}
Next, we present the outcomes of the WG activities: the identified framework for assessing LLM applications, and the ranking of LLM applications. Then, we comment on the technical and regulatory feasibility of these applications.

\subsection{Working Group Outcomes}

\paragraph{Vision and key evaluation criteria}
Based on the outcomes of the WG, we identified five major areas of improvement at the hospital: \emph{quality of care, patient engagement and satisfaction, administrative burden,  research and digital innovation, and continuous education.} Through discussions, brainstorming, and literature review, we then identified specific use-cases for LLMs applications that target these areas. See the complete list in \cref{app:survey-details}.

Moreover, during the discussions, the WG identified the following key evaluation criteria for applications: \emph{institutional impact}, \emph{impact on patient management and quality of care}, \emph{impact on patient satisfaction and engagement}, \emph{whether the impact is measurable}, \emph{whether the intended use is as software as medical device}, and \emph{safety}, defined as whether the risk of harm is low.

\paragraph{Application ranking} After two phases of application selection, the WG ranked the final seven use-cases. We present the average ranks, both across the evaluation criteria, and in total, in \cref{tab:use-cases}. We obtained two use-cases with the highest average ranking: First, the development of automatic tools for classifying and analyzing incident reports, e.g., cases of falls or incorrect drug dosage. Second, the summarization of discharge documentation into patient-adapted language. Both applications are of administrative nature, are deemed as easy to measure in terms of their impact, have benefits to patient engagement, and are assessed as being the least risky among the final application roster. 

Interestingly, the intermediate results from phase I showed the use-case of the assistive chatbot for clinical decision-making as both the most popular and best-ranked (see \cref{app:survey-details}). In phase II, however, after a detailed group discussion of all applications, it scored lower in the final anonymous survey, being ranked the worst in terms of safety, yet having the widest institutional impact.

\subsection{Technical and Regulatory Feasibility}

\paragraph{Regulation of medical uses}
Under the United States Food and Drugs Administration (FDA) regulations, European Law on Medical Devices Regulations (MDR), and the recently approved European Union (EU) AI Act, general purpose LLMs would not automatically be classified as medical devices \cite{meskoregulations}. It is the intended use that dictates the regulatory framework. As such, LLMs, or general-purpose LLMs, that are developed, fine-tuned or modified in order to serve a more specific medical purpose might be treated as medical devices. Classification as medical device triggers a series of requirements that may be challenging to meet for LLMs. For example, regulatory requirements apply to the entire development lifecycle of the device, not only to the phase where the model is adapted to medical purpose, posing a challenge for models that derive from general-purpose LLM. For these models, the development process most likely did not adhere to stringent medical device regulations. Even if it did, the resulting documentation may not be publicly accessible. 

Moreover, current risk assessment approaches may be inadequate for LLMs. For example, in the case of an \emph{assistive chatbot for clinical decision-making}---identified as a top priority use-case in our survey---several questions arise: How can we conduct a comprehensive risk assessment considering that the questions that future users may ask are potentially infinite? what role does the context in which the tool is used play in relation to its safety? Should we consider user-training as the main risk mitigation strategy for medical LLMs? Finally, in case of a LLMs-driven adverse event, would \emph {for-cause} auditing be possible considering LLMs limited explainability?  To date, these remain open questions, and proposed risk mitigation strategies focus on extensive user training and consequent user liability. In practice, starting from the assumption that users of medical LLMs tools would be capable of identifying subtle issues in LLMs output. This framework could be particularly problematic for applications such as the \emph{chatbot for medical education}, where users may have limited abilities to question LLMs-generated content. Regarding clinical support applications, clinical investigations will play a crucial role in LLMs risk assessment, with some investigations already ongoing~\cite{andreoletti2024ongoing}.

The regulatory framework for continual fine-tuning, or retraining, is another crucial point. In January 2021, the FDA took a step toward addressing this issue by publishing its action plan to facilitate AI-/ML innovation~\cite{fda}. The proposition was to regulate AI-powered software as medical device (SaMD) throughout their lifespan, introducing the so-called \emph{``predetermined change control plan.''} The guiding principles for the predetermined change control plans for ML-enabled medical devices were later published in October 2023. With this action the FDA aimed at aligning the speed of regulatory certification with the rapid release of new highly-performant ML models, including LLMs. The EU followed a similar approach in the newly released AI Act \cite{AIAct}, which states that new certification is not deemed for changes in algorithms or algorithms performance that were pre-determined by the manufacturer and pre-assessed at the time of relevant conformity assessment. It remains unclear how this process will work for LLMs as further re-training or fine-tuning typically involve new, larger datasets and a significantly different architecture. Continuous re-training may be required for many of the discussed use-cases, given the continuous advancements of medical care, new guidelines and scientific discoveries.

\paragraph{Hallucinations and omissions}  In their outputs, LLMs tend to both provide contextually plausible but factually incorrect information, a phenomenon known as hallucinations, as well as potentially omit crucial pieces of information~\cite{ji2023survey}. Depending on the use-case, these could pose critical issues to the viability of solving a problem with LLMs. For instance, even though a \emph{chatbot for medical education} was deemed a useful tool in our survey, the working group members agreed that it poses significant risk of harm due to hallucinations and omission, thus potentially misleading inexperienced medical clinicians. Solving these issues is an open problem, and there is no consensus on whether the solution can exist at all, with hallucinations or omissions possibly being an inherent feature of the way the current generation of LLMs are produced~\cite{bender2021dangers}. The pragmatic solutions often require additional application-specific infrastructure on top of LLMs, such as retrieval-augmented generation~\cite{lewis2020retrieval} or precise tooling~\cite{schick2024toolformer}. The reliability and safety of medical use-cases where these are an issue will thus depend on the reliability of infrastructure for preventing hallucinations or incorrect omissions.

\paragraph{Bias} In certain ways, LLMs can be akin to parrots~\cite{bender2021dangers}, regurgitating low-quality information obtained from their training data, oftentimes containing text from web sites such as reddit and Wikipedia. It should come as no surprise that LLMs could propagate outdated or generally harmful medical beliefs rooted in pseudo-science, conspiracy theories, racism or sexism~\cite[see, e.g., ][]{omiye2023large}, and incorporate these beliefs in downstream use-cases, e.g., in the recommendations given in the \emph{chatbot} use-cases, be it for \emph{medical education} or \emph{assistance with clinical decision-making}, or \emph{entry notes summarization}. Like with hallucinations, it is unclear whether this issue is solvable with the current generation of LLMs, but pragmatic deployment requires rigorous evaluation in terms of bias~\cite{gallegos2024bias}.

\paragraph{Infrastructure and data protection} Although some medical institutions are able to partner with external LLM training and inference providers~\cite{umeton2024gpt}, this way of incorporating LLMs into clinical practice comes with issues. First, in the case of commercial providers, it leads to the political issue of ``opaque commercial interests''~\cite{toma2023generative} guiding healthcare solutions. Second, for use-cases involving patient data, e.g., \emph{summarization of discharge letters,} the usage of external infrastructure is complicated by the ethical and legal data protection responsibilities, especially in jurisdictions with strict data protection regulation such as the EU. Indeed, in such use-cases, personal data would have to be transferred outside of the hospital's computational infrastructure either for model training or inference. According to General Data Protection Regulation \cite{GDPR}, e.g., the institutional requirements of our hospital, the data has to be sufficiently de-identified in such transfers. In the absence of highly reliable automated tools for deidentifying clinical text, however, every piece of the deidentified text might need to undergo manual inspection for missed personal identifiers. For instance, to release the CheXpert Plus dataset~\cite{chambon2024chexpert} of annotated chest X-ray images, up to 30 human annotators had to review more than \num{850000} text fragments. Thus, in the absence of tools for ensuring reliable deidentification, the usage of external infrastructure might require costly manual inspection of all of the patient-related data transferred to external providers.

\paragraph{On-premise infastructure costs} As we detailed previously, in jurisdictions with strict data protection regulation, outsourcing LLMs to external infrastructure is challenging for use-cases involving patient data. If we want to use LLMs for such use-cases, another option would be developing internal computational infrastructure on the medical institution's premises. At the same time, the state-of-the-art LLMs with multiple billions of parameters are notoriously costly not only to train, but even to infer from. For instance, a 65B parameter model might require four NVidia A100 GPUs each with 80GB for inference only~\cite{samsi2023words}. At the time of writing, such a setup would cost at least \num{60000} USD, in addition to recurring energy and personnel costs associated with keeping the infrastructure running. This leaves the question: Is the cost-benefit ratio worth it? Any LLM use-case should be rigorously evaluated in terms of the benefit it brings in terms of healthcare outcomes, education, or relieving administrative burden, against the infrastructure and associated costs.

\paragraph{Leakage of private information} Even if a model has been trained on-premise, LLMs can leak privacy-sensitive information contained in their training data~\cite{carlini2023quantifying}. This means that LLMs themselves as well as their outputs could contain personal information. In use-cases such as the \emph{assistive chatbot for clinical decision-making}, this means that if the model was trained on internal patient data, patient information could be extracted by malicious users, or could be leaked even in non-malicious chatbot interactions. There exists a formal theory of ensuring privacy in statistical and machine learning called \emph{differential privacy}~\cite{dwork2014algorithmic}, which enables to obtain privacy guarantees when models are trained on private data, preventing the leakage scenarios mentioned before. Recent work on fine-tuning language models with differential privacy~\cite{yu2022differentially} have shown promising results, even though normally differential privacy significantly reduces the utility of models. It is still, however, an open question, whether it is possible to obtain a meaningful level of privacy while retaining acceptable performance in high-risk downstream tasks such as assistive medical chatbots. 

\section{Discussion}
Our findings reveal a practical reality: although healthcare professionals see multiple promising applications for large language models in their practice, regulatory and infrastructural constraints significantly limit immediate implementation options. The preference for administrative applications, particularly discharge note generation and summarization, suggests a pragmatic path forward that balances regulatory compliance with feasibility.

Two key limitations of our study warrant discussion. First, the conclusions of our participatory methodology, which are limited to the specific context of Lausanne University Hospital, may not generalize across healthcare institutions. That said, we do hypothesize that our findings could largely generalize to similar medium-sized academic medical centers in Western Europe. Moreover, our study provides a framework for institutions to examine their unique needs and constraints through structured stakeholder engagement. Second, despite achieving representation across departments and roles, our participant pool likely shows selection bias, as the self-selected nature of participation in the working group means we may have captured perspectives from professionals who are inherently more enthusiastic about language models in healthcare.

Despite these limitations, our findings highlight how successful implementation of generative models in healthcare settings requires careful navigation of practical constraints, particularly around medical device regulation and data protection.

\bibliographystyle{unsrtnat}
\bibliography{main}

\newpage

\appendix

\section{Phase I: Survey}
\label{app:survey-details}

\paragraph{Methods}

In the survey, we allowing the participants to vote for the five most relevant use-cases from the pre-defined list and rank them according to the identified evaluation criteria. The survey was developed in Qualtrics Version, [06.2024].

\paragraph{Results}
Sixteen members of the WG, corresponding to one representative per hospital service, participated in the survey. 

We show the ranking of LLM use-cases by popularity in \cref{tab:phase1-use-cases}. \emph{Assistive chatbot for clinical decision-making}, \emph{chatbot for medical education} and \emph{automatic generation of discharge letters} were the most voted use-cases with 10, 8 and 7 votes respectively.

To scope the WG activities, for the next phase of the assessment, we only proceeded with the use-cases which were voted by at least five WG members. These are the top 8 use-cases in \cref{tab:phase1-use-cases}. To further simplify the discussion, we merged the following applications: smart retrieval of guidelines, chatbot for medical education, and smart summarization of current literature, resulting in six use-cases. During the following discussions, WG members proposed another use-case of classification and analysis of incident reports, which was agreed to be added to the final list of applications for the next phase, resulting in seven total use-cases.

\begin{table}[t]
\newcommand{\rankcolor}[2]{%
  \ifnum#1=1\textcolor[HTML]{472877}{#2}\else%
  \ifnum#1=2\textcolor[HTML]{3E4989}{#2}\else%
  \ifnum#1=3\textcolor[HTML]{30688E}{#2}\else%
  \ifnum#1=4\textcolor[HTML]{25838E}{#2}\else%
  \ifnum#1=5\textcolor[HTML]{1E9E89}{#2}\else%
  \ifnum#1=6\textcolor[HTML]{35B778}{#2}\else%
  \ifnum#1=7\textcolor[HTML]{6ECE58}{#2}\else%
  \ifnum#1=8\textcolor[HTML]{B5DE2B}{#2}\else%
  \textcolor[HTML]{000000}{#2}\fi\fi\fi\fi\fi\fi\fi\fi%
}
\setlength{\arrayrulewidth}{0.5pt}
\arrayrulecolor{lightgray}
\centering
\vspace{1em} %
\caption{Use-cases considered by the working group in Phase I (survey).}
\vspace{1em} %
\label{tab:phase1-use-cases}
\resizebox{\linewidth}{!}{
\begin{tabular}{l *5{w{c}{.5cm}|} c}
 & \multicolumn{5}{c|}{Areas} & \multicolumn{1}{c}{Ranking} \\ 
\textbf{Applications} & \textbf{QC} & \textbf{RI} & \textbf{PE} & \textbf{Adm} & \textbf{Edu} & \textbf{Popularity} \\
\cmidrule(l{3pt}r{3pt}){1-1} \cmidrule(l{3pt}r{3pt}){2-6} \cmidrule(l{3pt}r{3pt}){7-7}
\rowcolor[HTML]{F7F6F6}
Assistive chatbot for clinical decision-making & \yep & \yep & & & & \rankcolor{1}{1} \\
Chatbot for medical education & & & & & \yep & \rankcolor{2}{2} \\
\rowcolor[HTML]{F7F6F6}
Automatic generation of draft discharge letters & & & & \yep & & \rankcolor{3}{3} \\
Summarization of discharge letters into patient-adapted language & \yep & & \yep & & & \rankcolor{4}{4} \\
\rowcolor[HTML]{F7F6F6}
Analysis of patients' feedback and complaints & \yep & & \yep & & & \rankcolor{5}{5} \\
Smart retrieval of guidelines & \yep & & & & \yep & \rankcolor{5}{5} \\
\rowcolor[HTML]{F7F6F6}
Entry notes and medical records summarization & \yep & & & \yep & & \rankcolor{5}{5} \\
Smart summarization of current literature & & & & & \yep & \rankcolor{5}{5} \\
\rowcolor[HTML]{F7F6F6}
Smart search over patients records & \yep & \yep & & & & \rankcolor{6}{6} \\
Chatbot to increase health literacy in patients & & & \yep & & & \rankcolor{6}{6} \\
\rowcolor[HTML]{F7F6F6}
Clinical trial matching & & \yep & & & & \rankcolor{7}{7} \\
Automatic generation of imaging reports & \yep & & & \yep & & \rankcolor{7}{7}
\end{tabular}
}
\\[1em]
\scriptsize
\textbf{QC}: Quality of care, \ 
\textbf{RI}: Research \& innovation, \
\textbf{PE}: Patient engagement \& satisfaction, \ 
\textbf{Adm}: Administrative burden, \ 
\textbf{Edu}: Education \ 
\end{table}

\end{document}